\documentclass{iaus}
\usepackage{graphicx}

\title[N$_H$ variability in AGNs] 
{The structure of AGNs from X-ray absorption variability}

\author[G. Risaliti]   
{Guido Risaliti$^{1,2}$}

\affiliation{$^1$ INAF - Osservatorio Astrofisico di Arcetri, Largo E. Fermi 5, 
Firenze, Italy\\ email: {\tt risaliti@arcetri.astro.it} \\[\affilskip]
$^2$Harvard-Smithsonian Center for Astrophysics, 60 Garden Street, Cambridge, MA 02138, USA}

\pubyear{2009}
\volume{267}  
\pagerange{119--126}
\jname{Co-Evolution of Central Black Holes and Galaxies}
\editors{B.M.\ Peterson, R.S.\ Somerville, \& T.\ Storchi-Bergmann, eds.}
\begin{document}

\maketitle

\begin{abstract}
We present new evidence of X-ray absorption variability on time scales from a few hours to a few days for several nearby bright AGNs. The observed N$_H$ variations imply that the X-ray absorber is made of clouds eclipsing the X-ray source with velocities in excess of 10$^3$~km~s$^{-1}$, and densities, sizes and distances from the central black hole typical of BLR clouds. 
We conclude that the variable X-ray absorption is due to the same clouds emitting the broad emission lines in the optical/UV. We then concentrate on the two highest signal-to-noise spectra of eclipses, discovered in two long observations of NGC~1365 and Mrk~766, and we show that the obscuring clouds have a cometary shape, with a high density head followed by a tail with decreasing $N_H$. Our results show that X-ray time resolved spectroscopy can be a powerful way to directly  measure the physical and geometrical properties of BLR clouds.


\end{abstract}

\firstsection 
\section{Introduction}
X-ray absorption variability is a common feature in Active Galactic Nuclei (AGN).
An analysis of a sample of nearby X-ray obscured AGN with multiple X-ray
observations, performed a few years ago (Risaliti et al.~2002) revealed that column density (N$_H$) variations are almost ubiquitous
in local Seyfert galaxies. More recent observations performed with {\em XMM-Newton}, {\em Chandra} and {\em Suzaku} further confirmed this finding. The physical implications of these measurements 
are that the 
circumnuclear X-ray absorber (or, at least, one component of it) must be clumpy, and located at
sub-parsec distances from the central source. 
The comparison between different observations, typically performed at time distances of months-years, only provides upper limits to the intrinsic time scales of $N_H$ variations. An improvement of 
these estimates can only be obtained through observational campaigns within a few weeks/days, and/or
through the search for N$_H$ variations within single long observations.
Such short time-scale studies have been already performed for a handful sources: NGC~1365 (Risaliti et al.~2005, 2007, 2009), NGC~4388 (Elvis et al.~2004), NGC~4151 (Puccetti et al.~2007). In particular,
in the case of the AGN in NGC~1365 we revealed extreme spectral changes, from Compton-thin (N$_H$ in
the range 10$^{23}$~cm$^{-2}$) to reflection-dominated (N$_H$$>$10$^{24}$~cm$^{-2}$) in time
scales from a couple of days to $\sim$10~hours. 
Such rapid events imply that the absorption is due to clouds with velocity v$>$10$^3$~km~s$^{-1}$, at
distances of the order of 10$^4$ gravitational radii (assuming that they are moving with Keplerian velocity around the central black hole). 
The physical size and density of the clouds are of the order of 10$^{13}$~cm and 10$^{10}$-10$^{11}$~cm$^{-3}$, respectively.\footnote{For a detailed derivation and discussion of these parameters, we 
refer to Risaliti et al.~2009. A rough estimate is also mentioned here in Section~2.} All these physical parameters are typical for Broad Line Region (BLR) clouds, strongly suggesting that the X-ray absorber and the clouds responsible for broad emission lines in the optical/UV are one and the same.

Here we present more evidence, based on time resolved X-ray spectroscopy, of variable X-ray absorption due to BLR clouds, and we show how in the best cases we can use X-ray observations to directly probe the nature and  physical state of BLR clouds.

\section{X-ray absorption from BLR clouds in AGNs}

\begin{figure}
\begin{center}
 \includegraphics[width=13.5cm]{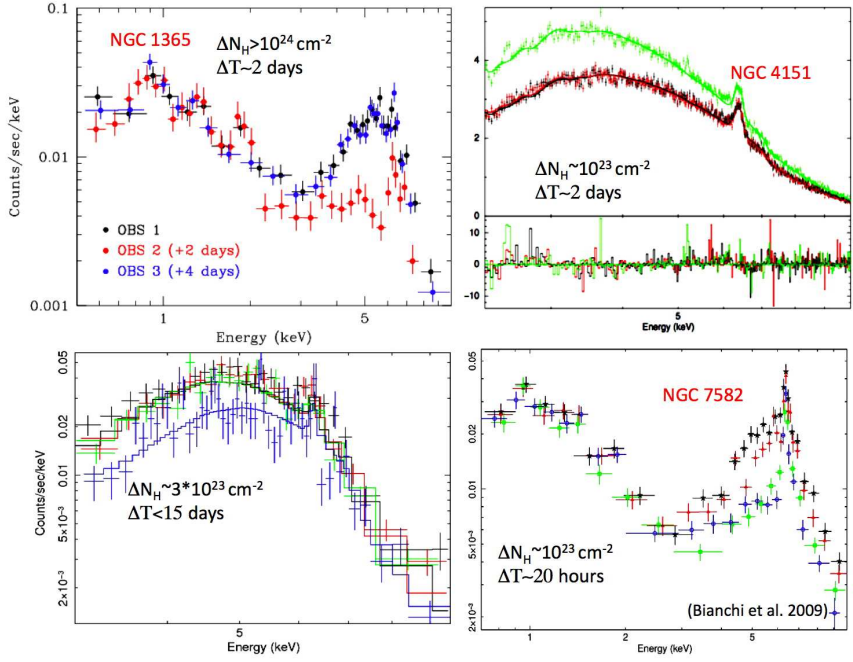} 
 \caption{Four examples of new N$_H$ variations on time scales of a few days discovered with campaigns 
of snapshot observations.}
\label{fig3}
\end{center}
\end{figure}

X-ray absorption due to BLR clouds crossing the line of sight to the observer implies variability time scales from hours to weeks, depending of the physical size of the X-ray source. We can obtain a rough estimate of the occultation times assuming a cloud moving with Keplerian velocity around the central black hole, and a size of the X-ray source of 10~R$_G$: T$\sim$3$\times$10$^5$~M$_7$~v$_3^{-1}$, where
M$_7$=M/(10$^7$~M$_\odot$) and v$_3$=v/(10$^3$~km~s$^{-1}$). 
Two observational methods have been used to investigate $N_H$ variations at these time scales:\\
1) Snapshot observations of the same source, 
repeated every few days.
After the first successful campaign of six short (10~ks) {\em Chandra} observations of NGC~1365,
we repeated this approach for other bright Compton-thin sources with column densities between 10$^{23}$ and 10$^{24}$~cm$^{-2}$, using both archival data and dedicated observations. In this way we
discovered clear cases of $N_H$ variations in NGC~4151 and UGC~4203 (one of the few "changing look" sources, known for having been observed in the past in both Compton-thin and reflection-dominated states). A further case of such variations has been discovered by Bianchi et al.~(2009) in NGC~7582 (Fig.~1). \\
2) Study of the spectral variations of bright sources during long observations.
This method consists of a two-phase analysis: we first use the hardness-ratio light curve to 
select the time intervals where strong spectral variations occurred; we then perform a complete
analysis of the spectra obtained from these intervals, in order to measure possible N$_H$ varations
(and to check if the spectral changes are due to other effects, such as variations of the slope of the continuum emission).

\begin{figure}
\begin{center}
 \includegraphics[width=13.0cm]{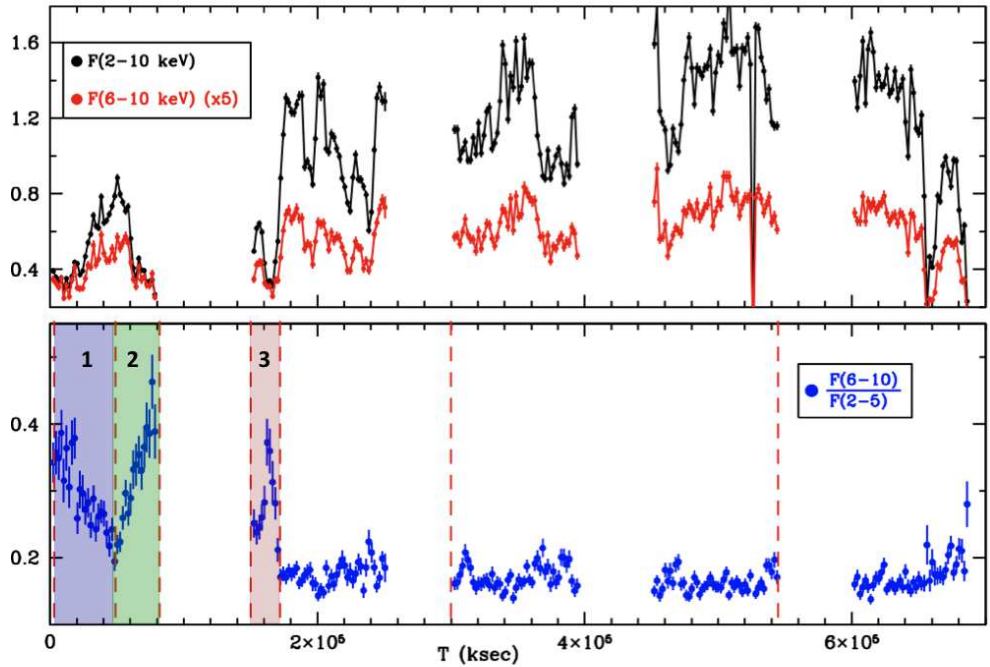} 
 \caption{Flux (top) and hardness ratio (bottom) light curves from the {\em XMM-Newton} long observation of Mrk~766. The observation is made in five consecutive {\em XMM-Newton} orbits. }
   \label{fig3}
\end{center}
\end{figure}
This approach is illustrated in Fig.~2 for the long {\em XMM-Newton} observation of Mrk~766. 
A complete analysis of these data is presented in Turner et al.~2007 and Miller et al.~2007. With respect to these works, our analysis is in many aspects less detailed (though the results are in full agreement), but is more effective in isolating the effects of possible N$_H$ variations. We note that Mrk~766 is a Narrow Line Seyfert~1, so on average we do not expect to observe complete X-ray 
absorption of the X-ray source. However, as we show below, isolated clouds occasionally cross the line of sight, producing measurable absorption in the X-ray spectrum. 
The upper panel of Fig.~2 shows the standard 2-10~keV flux  light curve for this observation, with the well known strong variability on time scales of thousands of seconds, or even shorter. 
The lower panel shows the light curve of the (6-10~keV)/(2-5~keV) flux ratio. In general, this light curve shows much smaller variations, indicating that the continuum shape remains the same during most of the luminosity variations. However, clear exceptions are observed in at least three intervals, 
highlighted in Fig.~2. During these intervals  it is possible that a cloud with N$_H$ of the order of 10$^{23}$~cm$^{-2}$ has covered the central source, strongly decreasing the observed flux in the soft band, without affecting the hard band, and therefore increasing the observed hardness ratio.

\begin{figure}
\begin{center}
 \includegraphics[width=11.5cm]{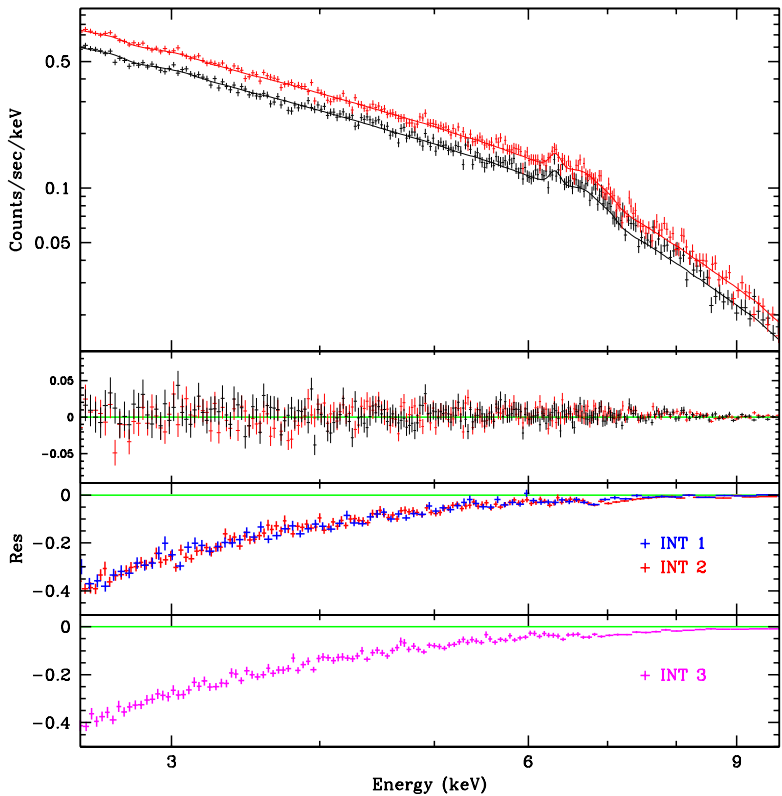} 
 \caption{Results from the spectral analysis of the eclipses observed in Mrk~766. Top two panels: spectra, best fit model and residuals from the third and fourh orbit (Fig.~2), where no spectral changes are observed. Bottom panels: difference between the spectra in the three intervals with spectral variations (Fig.~2) and the 
best fit model for the third and fouth orbit.} 
\label{fig3}
\end{center}
\end{figure}

Following this interpretation, in the first interval we should be observing  a cloud uncovering the X-ray source (with the covering phase occurred before the beginning of the observation); in the second interval another cloud shoud be covering the X-ray source, with the uncovering phase not observed due to the "dead time" between two consecutive {\em XMM-Newton} orbits. Finally, in the third interval we should be observing the whole eclipse. 
\begin{figure}
\begin{center}
 \includegraphics[width=13.5cm]{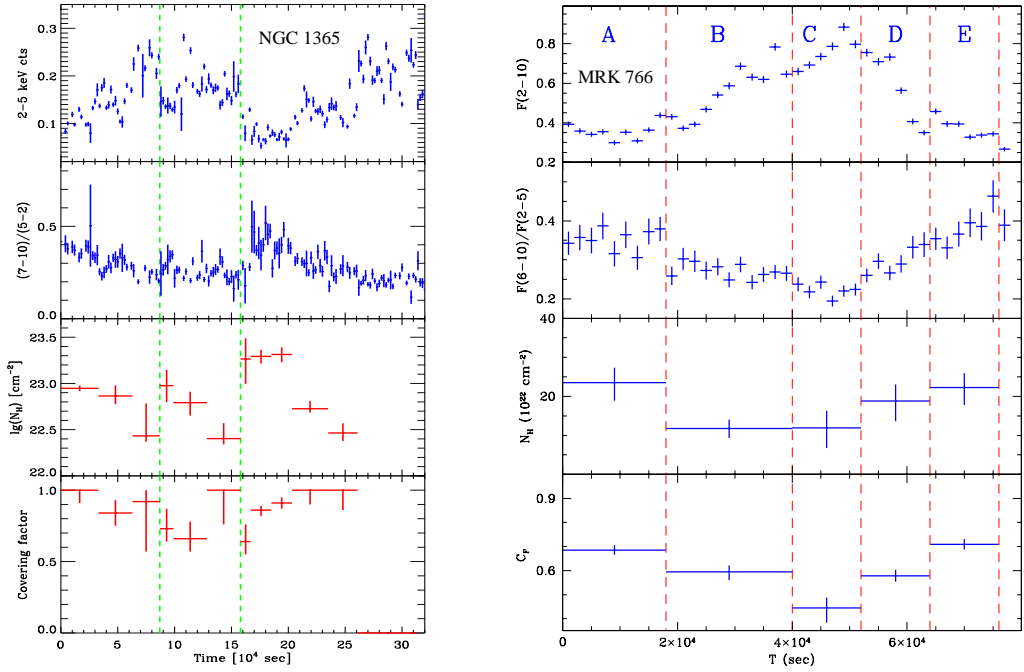} 
 \caption{Detailed analysis of a {\em Suzaku} observation of NGC~1365 (left panel) and of the first
two intervals of the Mrk~766 observation highlighted in Fig.~2 (right panel). In each panel, we show, from top to bottom, the light curves for: flux, hardness ratio (in the case of Mrk~766 these two plots are just a zoom of the light curves shon in Fig.~2), N$_H$, and  covering factor. }
   \label{fig3}
\end{center}
\end{figure}

In order to check this scenario, we performed a complete analysis of the spectra obtained from the
three highlighted intervals, and of those obtained from the third and fourth orbit, representing the standard spectral state of the source.  
In this analysis we allowed all the main spectral parameters  
of the model to vary among the different intervals. The results of the this study, illustrated in Fig.~3 are the following:
1) the 2-10~keV spectrum obtained from the third and fourth orbit (the "standard" state) is well reproduced by a typical model for type~1 AGNs, consisting of a power law, a reflection component and an iron emission line; 2) the spectral variations observed in the three intervals discussed above are completely reproduced by three absorption components with column densities in the range 1-3~10$^{23}$~cm$^{-2}$.
\begin{figure}
\begin{center}
 \includegraphics[width=13.5cm]{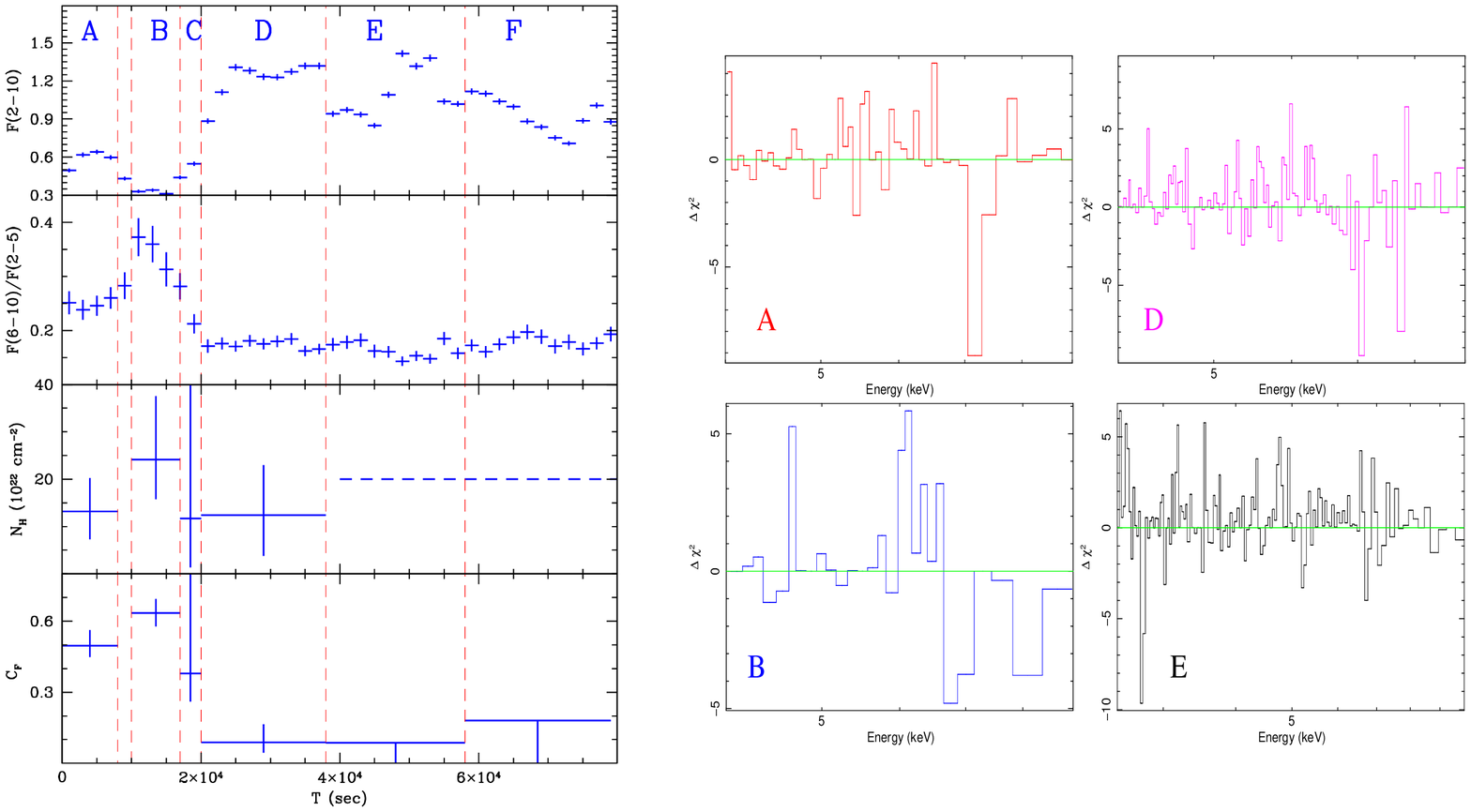} 
 \caption{Left panel: same as in Fig.~3, for the second orbit of the {\em XMM-Newton} observation of Mrk~766. Right panel: residuals for the intervals indicated in the left panel. The presence of iron absorption lines suggest the presence of a ionized tail. }
   \label{fig4}
\end{center}
\end{figure}

The analysis briefly summarized here has been applied to several sources with long archival observations by {\em XMM-Newton} and {\em Suzaku}. Up to now, we obtained unambiguous evidence for $N_H$ variations in
NGC~4388, MCG-6-30-15, Mrk~766, NGC~1365. A further case of similar variations has been found in a {\em Suzaku} observation of NGC~3227 (Liu et al.~2010, subm.).

A summary of all the positive detections obtained so far is shown in Table~1. In all cases the inferred
physical parameters for the obscuring clouds clearly indicate that the X-ray absorption is due to BLR clouds.
\section{Structure of the BLR clouds from X-ray spectroscopy}

The analysis presented above can be further refined in the few cases where the statistics is high enough to perform a more detailed study, or where multiple occultation event occur.
In particular, we briefly summarize two results from the analysis of the two so-far best studied sources, Mrk~766 and NGC~1365.\\
{\bf 1) Velocity distribution of BLR clouds.} The analysis of the spectral variations in Mrk~766 (Fig.~2-4) directly provides contraints on the distribution of BLR cloud velocities, from the comparison of the durations of the observed occultations. 
The actual values of the cloud velocity in each case cannot be precisely determined, being dependent on the exact size of the X-ray source. However, the ratio between the durations of the different eclipses is an estimate of the velocity ratios. In particular, the duration of the two first occultation events (Fig.~2), is about 4-5 times longer than that of the third eclipse. This suggests a similar spread in the velocity of BLR clouds. \\
{\bf 2) Structure of BLR clouds.} The model adopted to reproduce the observed occultation events consists of an absorption component with a constant (during the eclipse) N$_H$, and a variable covering fraction, $C_F$. This is an extreme  simplification of the cloud structure, which is assumed to have constant column density, and very sharp edges. The model can be improved by releasing the assumption of constant N$_H$. Doing so, in most of the cases analyzed so far we found that it is impossible to  significantly constrain the values of N$_H$ and $C_F$ during the different phases of an eclipse, due to their strong degeneracy. In other words, it is only possible to study the evolution of one parameter, freezing the other one. 
The only exceptions found so far, i.e. spectra with enough signal-to-noise to analyze the evolution of the two parameters at the same time, are Mrk~766 and NGC~1365.

In Fig.~4A we show the results of a recent analysis of a {\em Suzaku} observation of NGC~1365 (Maiolino et al.~2010, subm), where the hardness ratio light curve clearly shows at least two possible occultation events. We performed a complete spectral analysis of several short intervals before, during and after the eclipses, and we obtained the results shown in the two lower panels of Fig.~4, showing the light curves of the covering factor and the column density. The most interesting results come from the second
eclipse: the covering factor $C_F$ increases during the occultation phase, with a constant N$_H$$\sim$3$\times$10$^{23}$~cm$^{-2}$. Then, in the subsequent phase, $C_F$ remains compatible with 1 (i.e. complete covering) while the column density decreases. This behaviour can only be explained with a "cometary shape" of the obscuring cloud, with a  high column density head, and a tail with decreasing N$_H$.

An analogous result has been obtained (though with smaller statistical significance) for the first two eclipses in Mrk~766 (Fig.~4B). The third eclipse of Mrk~766 is instead too fast to allow a detailed spectral analysis (the spectral counts in such short time intervals are not enough to break the degeneracy between $C_F$ and $N_H$, as shown in Fig.~5). However, a strong indication of a cometary tail comes from the residuals with respect to the best fit model during and after the eclipse, showing strong absorption lines due to Fe~XXV in outflow with a velocity of 10-15000 km~s$^{-1}$. These features completely disappear after $\sim$40~ks.  A complete analysis of these spectra, where we demonstrate the high statistical significance of the lines detections, is presented in a dedicated paper (Risaliti et al.~2010, subm.) The highly ionized component revealed by the absorption lines strongly suggest the presence of a ionized outflowing tail associated to the obscuring cloud. 

We note that this highly ionized component is not present in the first two occultation events in Mrk~766, but only in the third, much faster event. Even if one case is clearly not enough to build a complete model, 
the data suggest a scenario where the obscuring clouds are distributed in a large range of distances from the center (if we assume Keplerian velocity, the factor $\sim$4-5 of spread in velocities corresponds to a factor of $\sim$20 of spread in distances), and where the clouds closer to the center are also the more ionized ones. Thsi simple scheme is illustrated in Fig.~6.   
\begin{table}
  \begin{center}
  \caption{List of sources with N$_H$ variations on short time scales}
  \label{tab1}
 {\scriptsize
  \begin{tabular}{|l|c|c|c|c|}\hline
Name     & $\Delta$(N$_H$)$^a$ & $\Delta$(T)$^b$ & Method$^c$ & Ref. \\
\hline
NGC 1365 & $>$10$^{24}$        &  $<$ 2~days     & Snapshot   & 1,2 \\
NGC 1365 & 3$\times$10$^{23}$  &  10 hours       & Continuous & 3,4 \\
NGC 4388 & 2$\times$10$^{23}$  &  15 hours       & Continuous & 1 \\
NGC 4151 &  2$\times$10$^{23}$ &  20 hours       & Continuous & 5 \\
NGC 4151 & 10$^{23}$  &        $<$2 days         & Snapshot   & 1 \\
NGC 7582 &          10$^{23}$  &  20 hours       & Snapshot   & 6 \\
Mrk 766  &  3$\times$10$^{23}$ &  10 to 20 hours & Continuous & 1,7 \\
MCG-6-30-15&         10$^{23}$ & 10 hours       & Continuous & 1 \\
UGC 4203 & 3$\times$10$^{23}$  & $<$ 15 days     & Snapshot   & 1 \\
NGC 3227 & 7$\times$10$^{22}$   & 1 day         & Continuous & 8 \\
\hline
  \end{tabular}
  }
 \end{center}
\vspace{1mm}
 \scriptsize{Notes: $^a$: N$_H$ variations in cm$^{-2}$; $^{b}$ duration of the observed eclipse;
$^c$: Observational method: repeated snapshot observations, or analysis of long  continuous observations.
References: 1: this work; 2: Maiolino et al.~2010, subm.; 3: Risaliti et al.~2007; 4: Risaliti et al.~2009; 
5: Puccetti et al.~2007; 6: Bianchi et al.~2009; 7: Risaliti et al.~2010, subm.; 8: Liu et al.~2010, in prep.}

\end{table}

\section{Conclusions, and future work} 

We have presented new time-resolved spectral studies of several bright AGNs, showing column density variability on time scales for a few hours to a few days. The main results of our analysis are:\\
1) Fast (hours-days) column density variability is common among AGNs. This implies that the observed variable  X-ray absorption is due to clouds with velocities, densities, sizes and distances from the central black hole of the same order of those of BLR clouds. \\
2) In the highest signal-to-noise studies, it is possible to investigate the structure and shape of the single obscuring clouds in detail. This reveals a "cometary" profile, with a high column density head and a tail with decreasing N$_H$.\\

These results show that the X-ray absorption is at least in part due to BLR clouds, and that X-ray spectroscopy can be a powerful tool to diretly measure the physical properties of the broad line region in AGNs.\\
\begin{figure}
\begin{center}
 \includegraphics[width=13.5cm]{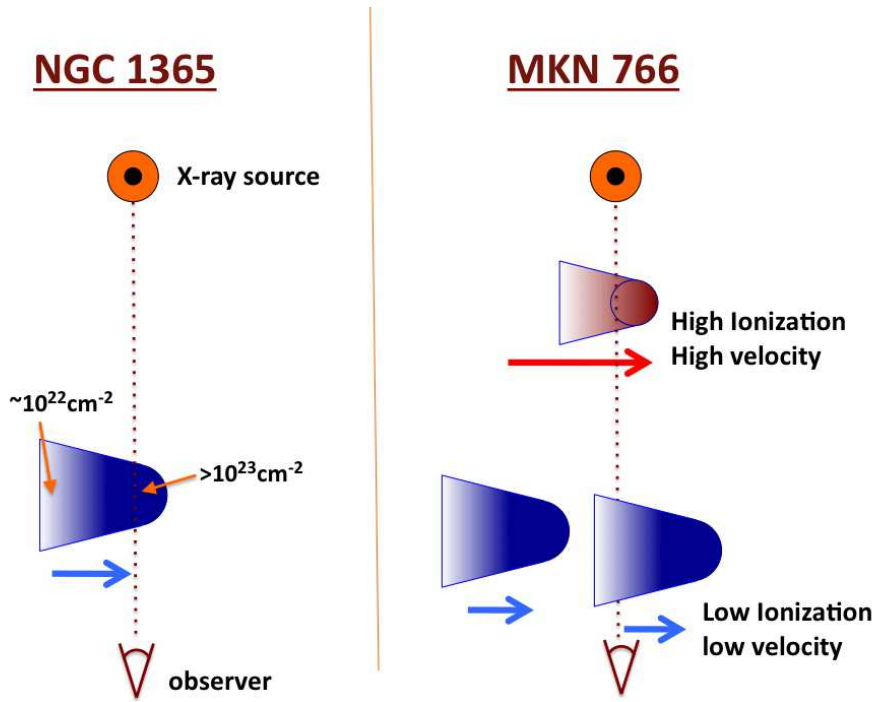} 
 \caption{A schematic view of the X-ray absorbing clouds, as estimated from the observations descrived here.
The eclipsing cloud in NGC~1365 is the one with the best-measured properties, with a column density going from $\sim$3$\times$10$^{23}$~cm$^{-2}$ in the head, to a few 10$^{22}$~cm$^{-2}$ in the tail (Maiolino et al.~2010). In Mrk~766, we suggest that the fast cloud with the ionized tail is much closer to the center than the two slower, neautral clouds. The ionized tail in the fast cloud is probably in outflow (not shown here; a complete analysis is presented in Risaliti et al.~2010).}
   \label{fig1}
\end{center}
\end{figure}

The work presented here is only the first part of an on-going comprehensive analysis of all the bright AGNs with long high quality X-ray observations. At present, the evidence of "common" 
short time scale N$_H$ variations presented here is based on a sparse sample of sources, with no homogeneous selection criteria. The next major step is therefore the selection and the homogeneous analysis of a representative sample of the local AGN population. This work will provide a quantitative estimate of the occurrence of N$_H$ variability on short time scales, and hopefully, will lead to the discovery of more high signal to noise spectra of occultations, such as the ones found in NGC~1365 and Mrk~766, in order to perform other studied of the physical propoerties of BLR clouds.\\

{\bf Acknowledgements.}. 
This work has been partially supported by NASA grants NNX08AX78G and G08-9107X,
and by grant ASII-INAF I/088/06/0

\end{document}